\documentclass[a4paper]{report}

\usepackage[sectionbib,round,numbers]{natbib}
\usepackage[utf8]{inputenc}
\usepackage[T1]{fontenc}
\usepackage{RJournal}
\usepackage{amsmath,amssymb,array}
\usepackage{booktabs}
\usepackage{graphicx}%
\usepackage{multirow}%
\usepackage{amsmath,amssymb,amsfonts}%
\usepackage{amsthm}%
\usepackage{mathrsfs}%
\usepackage[title]{appendix}%
\usepackage{xcolor}%
\usepackage{fancyvrb}
\usepackage{textcomp}%
\usepackage{manyfoot}%
\usepackage{booktabs}%
\usepackage{algorithm}%
\usepackage{algorithmicx}%
\usepackage{algpseudocode}%
\usepackage{listings}%
\usepackage[normalem]{ulem}
\usepackage{color}
\usepackage{booktabs}
\usepackage{geometry} 
\usepackage{array}
\usepackage{tabularx}
\usepackage{longtable}
\usepackage{caption} 
\usepackage{listings}    
\usepackage{listings}

\lstdefinestyle{prettytable}{
    backgroundcolor=\color{gray!10}, 
    basicstyle=\ttfamily\footnotesize, 
    columns=fullflexible,            
    frame=single,                    
    framerule=0.5pt,                 
    rulecolor=\color{gray},          
    keywordstyle=\bfseries,          
    showstringspaces=false,          
    tabsize=4,                       
    literate={0}{{\textcolor{blue}{0}}}{1}%
             {1}{{\textcolor{blue}{1}}}{1}%
             {2}{{\textcolor{blue}{2}}}{1}%
             {3}{{\textcolor{blue}{3}}}{1}%
             {4}{{\textcolor{blue}{4}}}{1}%
             {5}{{\textcolor{blue}{5}}}{1}%
             {6}{{\textcolor{blue}{6}}}{1}%
             {7}{{\textcolor{blue}{7}}}{1}%
             {8}{{\textcolor{blue}{8}}}{1}%
             {9}{{\textcolor{blue}{9}}}{1}%
}

\usepackage{adjustbox}  
\usepackage{fancyvrb}   

\definecolor{dkgreen}{rgb}{0,0.6,0}
\definecolor{gray}{rgb}{0.5,0.5,0.5}
\definecolor{mauve}{rgb}{0.58,0,0.82}

\useunder{\uline}{\ul}{}

\lstdefinestyle{mystyle}{
    backgroundcolor=\color{white},   
    commentstyle=\color{green!60!black},
    keywordstyle=\color{blue},
    numberstyle=\tiny\color{gray},
    stringstyle=\color{orange},
    basicstyle=\ttfamily\footnotesize,
    breakatwhitespace=false,         
    breaklines=true,                 
    captionpos=b,                    
    keepspaces=true,                 
    numbers=left,                    
    numbersep=5pt,                  
    showspaces=false,                
    showstringspaces=false,
    showtabs=false,                  
    tabsize=2
}

\lstdefinestyle{mystyle_table}{
    basicstyle=\ttfamily\footnotesize, 
    breaklines=false,                    
    frame=single,                       
    columns=fullflexible,               
    keepspaces=true,                    
    backgroundcolor=\color{lightgray!20} 
}

\begin{document}

\sectionhead{CMSAC Reproducible Research}
\year{2024}
\month{Nov}








\begin{center}
    \LARGE \textbf{KabaddiPy: A package to enable access to Professional Kabaddi Data} \\[1em]
    \normalsize
    \begin{tabular}{@{}c c@{}}
        \textbf{Bhaskar Lalwani}\textsuperscript{$\dagger$} & \textbf{Aniruddha Mukherjee}\textsuperscript{$\dagger$}\textsuperscript{*} \\
        \small School of Computer Engineering & \small School of Computer Engineering \\
        \small Kalinga Institute of Industrial Technology & \small Kalinga Institute of Industrial Technology \\
        \small \texttt{2205460@kiit.ac.in} & \small \texttt{2205533@kiit.ac.in} \\
    \end{tabular}
\end{center}

\renewcommand\thefootnote{}

\footnotetext{$\dagger$ Both of the authors contributed equally.}
\footnotetext{* Corresponding author: Aniruddha Mukherjee (\texttt{23f1003186@ds.study.iitm.ac.in})}

\abstract{
Kabaddi, a contact team sport of Indian origin, has seen a dramatic rise in global popularity, highlighted by the upcoming Kabaddi World Cup in 2025 with over sixteen international teams participating,
alongside flourishing national leagues such as the Indian Pro Kabaddi League (230 million viewers)
and the British Kabaddi League. We present the first open-source Python module to make Kabaddi statistical data easily accessible from multiple scattered sources across the internet. The module was developed by systematically web-scraping and collecting team-wise, player-wise and match-by-match data. The data has been cleaned, organized, and categorized into team overviews and player
metrics, each filterable by season. The players are classified as raiders and defenders, with their best strategies for attacking, counter-attacking, and defending against different teams highlighted. Our module enables continuous monitoring of exponentially growing data streams, aiding researchers to quickly start building upon the data to answer critical questions, such as the impact of player inclusion/exclusion on team performance, scoring patterns against specific teams, and break down opponent gameplay. The data generated from Kabaddi tournaments has been sparsely used, and coaches and players rely heavily on intuition to make decisions and craft strategies. Our module can be utilized to build predictive models, craft uniquely strategic gameplays to target opponents and identify hidden correlations in the data. This open source module has the potential to increase time-efficiency, encourage analytical studies of Kabaddi gameplay and player dynamics and foster reproducible research. The data and code are publicly available: \url{https://github.com/kabaddiPy/kabaddiPy}
}

\vspace{-0.5cm}

\section{Introduction}

\vspace{-0.3cm}

Kabaddi is a fast-paced team-contact sport, that has been rapidly gaining international recognition and popularity. In a game that can be loosely described as a combination of rugby, American football and tag \cite{kabaddinytimes} two teams take turns sending a player, called the ``raider'' into the opponent's half with the goal of tagging as many defenders as possible and returning to their own side without being tackled. The raider must accomplish this in a single breath for an offense lasting 30 seconds, all the while chanting ``kabaddi'' (pronounced \texttt{kuh-bud-DEE}).

Similar to American football, where the offense aims to evade the opponent's defense to score touchdowns, the raider in Kabaddi must dodge the defender's tackles, tag them and return to their own half to score points. Meanwhile the defense must work in coordination, aiming to tackle and immobilize the raider before they can return. Unlike football, Kabaddi requires no ball or protective gear; it’s a minimalist sport that focuses purely on strategy and physical strength.



Originally a traditional Indian sport, Kabaddi was first exhibited at the 1936 Berlin Olympics~\citep{olym}. \textcolor{black}{Since becoming a regular feature in the Asian Games in 1990~\citep{asian-games}, the sport has gained international recognition, as witnessed by the launch of the Kabaddi World Cup in 2004, with England set to host the 2025 edition~\citep{bkl_1, kwc}.} The sport's popularity skyrocketed with the inception of the Indian Pro Kabaddi League (PKL) in 2014, now attracting 230 million viewers, second only to cricket's Indian Premier League (IPL).

With success of the PKL, several regional leagues, like the European Kabaddi Championship and the British Kabaddi League, have emerged and over 30 nations have established dedicated national Kabaddi teams including the United States~\citep{uskabaddi}, Japan, South Korea and Iran. 




For a game rooted in strategy, which has witnessed growing viewership across the board, and its highly commercial nature - Star Sports bought the media rights for PKL for \$120 million for five years~\citep{sponsorship}, Kabaddi is an overlooked sport. It still lacks the analytical infrastructure seen in other sports such as hockey, basketball or football~\citep{howell_gilani_fastRhockey_2021}. Existing sports analytics research in Kabaddi  has been hampered by easy access of publicly available data. The limited studies conducted have had high barriers of research, manually web scraping data from the official Pro Kabaddi League (PKL) website.~\citep{singh2023kabaddi_springer, parmarkabaddi, Khullar2024Mar}.

In this paper, we introduce the first open source module for aggregating Kabaddi data from the PKL website and other disparate sources. Through this module, we aim to improve access to Kabaddi data, standardize it into a single source, and contribute to the formation of a community of researchers and analysts around Kabaddi, increasing the potential for development of sophisticated strategies and detailed insights to improve team and player performance.

\vspace{-0.35cm}

\section{Overview of Kabaddi Play}

\vspace{-0.35cm}

Kabaddi is set on a rectangular court, measuring 10 by 13 metres (33 feet by 43 feet) for men and 8 by 12 metres (26 feet by 39 feet) for women, divided into two halves by a \textit{midline}. Two teams of seven players compete on opposite ends of the court. The game consists of two 20-minute halves separated by a 5-minute halftime break during which the teams switch sides.


The offense operates as a unique ``tag and return'' system~\citep{taylorandfran}, distinct from the ball-focused scoring of American football. The offensive player, the raider, sprints into the opponent’s half, seeking to tag one or more defenders, and return to their own half without being tackled, all in a single breath. A tag can be made using either a hand or a foot and each successful tag earns a point, with bonus points awarded for tagging multiple defenders or clearing the opponent’s court. This is known as a ``\textit{raid}'', and an entire raid must be completed in no more than 30 seconds.
It is crucial that this be done in one breath, hence the necessity to continuously chant ``kabaddi''.

While Kabaddi's defense shares similarities with that of rugby, the defenders focus on preventing the raider from returning to their own half, rather than keeping them out. If the defenders successfully tackle and hold the raider before they can return, the defending team earns a point. Additionally, the defense scores if the raider goes out of bounds or fails to return to their half before exhaling (when they stop chanting ``kabaddi''). 

If the defenders attempt a tackle but fail to prevent the raider from returning to their own half, all defenders involved in the tackle are considered tagged, resulting in points being conceded to the raiding team.
This creates a unique, high-stakes dynamic where every raid becomes a time-critical scoring opportunity for both the offense and defense. Players who are tagged or tackled are taken out of the game but can be ``\textit{revived}'' when their team scores from a successful tag or tackle.

\begin{figure}[h]
    \centering
    \includegraphics[width=0.6\linewidth]{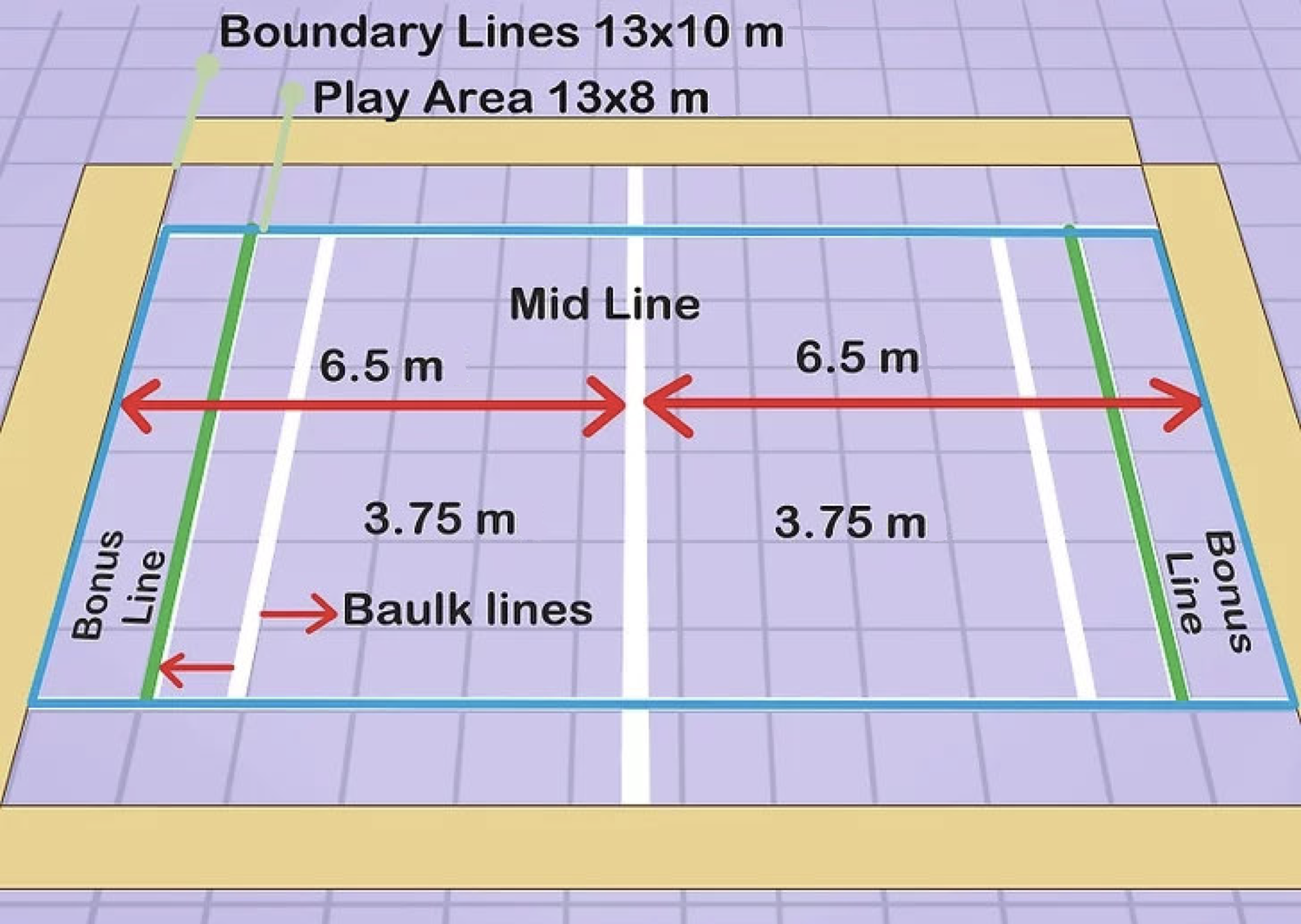}
    \caption{Standard Style Kabaddi Mat.~\cite{wikihow}}
    \label{fig:enter-label}
\end{figure}

\vspace{-0.1cm}

On the court, key lines play a crucial role in the point system. The \textit{baulk line}, which is 3.75 meters (12.3 feet) from the midline, acts as a marker that the raider must cross to make the raid legal; if they fail to cross it without making a tag, no points are scored, and the raid is considered unsuccessful. The \textit{bonus line}, located 1 meter (3.2 feet) beyond the baulk line, offers the raider an opportunity to score an additional point, if they cross it while keeping at least one foot in the air and manage to return without being tackled. This is only possible when there are six or more players on the mat.

In addition to standard gameplay, special situations such as \textit{do-or-die raids} and \textit{all-outs} significantly impact scoring dynamics.
A team can further increase their score by achieving an all-out, which involves eliminating all the opposing players on the court. This forces the opposition to reset their players on the court, and concede two points.
A do-or-die raid occurs, after two consecutive empty raids, and the team must score points in it or face a penalty of one point. The green, yellow, and red cards in Kabaddi are analogous to soccer's warning, temporary suspension, and ejection penalties. A green card is like a verbal warning, a yellow card mirrors a player being temporarily sidelined (e.g., a 2-minute suspension), and a red card resembles a player being ejected from the game. Receiving a yellow or red card awards one technical point to the opposing team.

The strategic depth of Kabaddi lies in how teams deploy their raiders and defenders, utilizing formations, timing, and coordination to gain an advantage. Defensive tactics, such as chain tackles and ``super tackles'', where a shorthanded team successfully tackles a raider, add another layer of complexity to the game.

\vspace{-0.1cm}

\noindent Various defensive and offensive strategies for scoring points are described briefly, in Table~\ref{tab:defense} and Table~\ref{tab:offense}~\citep{singh2023game}.

\vspace{-0.15cm}

\begin{table}[h]
\centering
\begin{minipage}{0.49\textwidth}
\resizebox{\textwidth}{!}{%
\begin{tabular}{|l|l|}
\hline
\multicolumn{1}{|c|}{\textbf{Defensive Move}} & \multicolumn{1}{c|}{\textbf{Description}}                                                                     \\ \hline
Waist/Back Hold & \begin{tabular}[c]{@{}l@{}}When the defender attempts to grab the raider\\ mid-air by the waist to pin them down on \\ the mat.\end{tabular}           \\ \hline
Ankle Hold                                    & \begin{tabular}[c]{@{}l@{}}The defenders attempt to stop the raider \\ by grabbing the ankle.\end{tabular}    \\ \hline
Thigh Hold                                    & \begin{tabular}[c]{@{}l@{}}The defenders attempt to hold the thigh of \\ raider with both hands.\end{tabular} \\ \hline
Block           & \begin{tabular}[c]{@{}l@{}}The defenders physically try to stop the raider \\ from crossing the mid-line and go back to their \\ court.\end{tabular}   \\ \hline
Chain Tackle    & \begin{tabular}[c]{@{}l@{}}When two or more defenders attempt a \\ co-ordinated tackle to prevent the raider \\ from crossing the midline\end{tabular} \\ \hline
Dash            & \begin{tabular}[c]{@{}l@{}}When the defender pushes the raider out \\ of the court by "dashing". Earns a point \\ for the defense.\end{tabular}        \\ \hline
\end{tabular}%
}
\caption{Table describing various defending strategies used in Kabaddi}
\label{tab:defense}
\end{minipage}%
\hfill
\begin{minipage}{0.49\textwidth}
\centering
\resizebox{\textwidth}{!}{%
\begin{tabular}{|l|l|}
\hline
\multicolumn{1}{|c|}{\textbf{Offensive Move}} & \multicolumn{1}{c|}{\textbf{Description}}                                                                      \\ \hline
Toe Touch                                     & \begin{tabular}[c]{@{}l@{}}The raider attempts to score by touching\\ just the defenders' toe.\end{tabular}    \\ \hline
Hand Touch                                    & \begin{tabular}[c]{@{}l@{}}Raider attempts to score by touching \\ the defenders' with their arm.\end{tabular} \\ \hline
Front and Side Kick                           & \begin{tabular}[c]{@{}l@{}}The raider attempts to score by kicking \\ in front or sideways.\end{tabular}       \\ \hline
Reverse/Back Kick & \begin{tabular}[c]{@{}l@{}}Raider can turn their back towards the \\ defender and kick backwards to score a \\ touch point.\end{tabular}               \\ \hline
Leg Thrust        & \begin{tabular}[c]{@{}l@{}}The raider uses their leg strength to \\ push through the defenders; used \\ when trapped in a hold or tackle.\end{tabular} \\ \hline
Dubki (duck)      & \begin{tabular}[c]{@{}l@{}}The raider ducks below the defenders \\ to reach the half line. Used to avoid chain \\ tackle\end{tabular}                  \\ \hline
\end{tabular}%
}
\caption{Table with diverse raider strategies used in Kabaddi}
\label{tab:offense}
\end{minipage}
\end{table}

\vspace{-0.2cm}

\section{Methodology/Data Collection}

\vspace{-0.2cm}

\texttt{KabaddiPy} has been developed as a comprehensive Python package that has aggregated historical data over 10 previous Pro Kabaddi League (PKL) seasons, through web scraping multiple sources: the official PKL website, ProKabaddi.com~\citep{prokabaddicom}, Tableau Dashboard prepared by SportsKPI \citep{tableau_kpi}, kabaddiadda.com~\citep{kabaddiadda} and Global Sports Data Archive~\citep{gsa-archive}. Play-by-play and player statistics were primarily sourced from ProKabaddi.com, while team-level data was gathered from the Tableau Dashboard. Partial auction data was scraped from kabaddiadda. Additional metrics for raider performance were scraped from the Tableau Dashboard. The data was cross-verified with information from the official PKL website to ensure its validity.


The scraper populated a central repository with the historical data (\href{https://github.com/kabaddiPy/kabaddi-data}{github-kabaddi-data})  for ease of access. The repository will be constantly updated with live data when the PKL season is going on.

In addition to these core functions, the package offers several helper functions specifically designed to process, parse and clean the raw data into structured formats. While these processing functions were developed as part of the Kabaddi package, they are not expected to be used by anyone for Kabaddi analytics. However, to ensure full reproducibility and transparency, we provide both the web scraping functions and the collated data. This allows researchers to not only access pre-processed data (by function calls) but also to replicate the data collection process.

\vspace{-0.55cm}

\section{Installing KabaddiPy}

\vspace{-0.25cm}

KabaddiPy, (v1.0.0) is available on the \href{https://pypi.org/project/kabaddiPy/}{Python Package Index} (PyPI) and can be downloaded using pip.

\begin{lstlisting}
pip install kabaddiPy
\end{lstlisting}

\vspace{-0.2cm}
\noindent The class can be initialized with the below. All the functions belong to this class and can be accessed accordingly.

\begin{lstlisting}
import kabaddiPy

# Initialize the aggregator
pkl = kabaddiPy.PKL()
\end{lstlisting}

\vspace{-0.67cm}

\section{Module Usage}

\vspace{-0.1cm}

\texttt{kabaddiPy} enables an analyst to start from very basic data, such as the season standings, and move to advanced statistics such as, the effectiveness of raiders against a given number of defenders for a given team or zone-wise strength of players.

The module makes understanding the sport for those new to it (a large portion of our audience) easy, while still allowing for experts to conduct an in-depth analysis of strategies at the team, match or player level.

To demonstrate the functionality of \texttt{kabaddiPy}, we identify a pivotal match from last season (season 10): the game that secured Puneri Paltan's place in the final, ultimately leading to their championship victory.

\begin{lstlisting}
matches = pkl.get_season_matches(season=10)

result = matches[(matches['League_Stage'] == 'Semi Final') & 
 ((matches['team_name_1'] == 'Puneri Paltan') | (matches['team_name_2'] == 'Puneri Paltan'))]

# selecting specific columns to display
print(result[['Season','Match_ID','League_Stage','Match_Outcome',
'team_score_1','team_score_2','team_name_1','team_id_1','team_name_2','team_id_2', 'Winning Margin']])

    
\end{lstlisting}

\begin{lstlisting}[style=mystyle_table]
Season  Match_ID  League_Stage  Match_Outcome     team_    team_   team_    team_  team_     team_  Winning 
                                                  score_1  score_2 name_1   id_1   name_2    id_2   Margin
---------------------------------------------------------------------------------------------------------------
10      3163      Semi Final    Puneri Paltan     37       21      Puneri   7      Patna     6      16
                                won by 16 Pts                      Paltan          Pirates
#
#...with 6 additional columns 'Match_Name', 'Year', 'Venue','Start_Date', 'End_Date' and 'Result', 
\end{lstlisting}

Crucially, we are able to get the unique \texttt{Match\_ID} here. We can use this \texttt{ID} to query functions to know exactly why Puneri Paltan won the match by a huge margin.

\begin{figure}[h]
    \centering
    \includegraphics[width=0.85\linewidth]{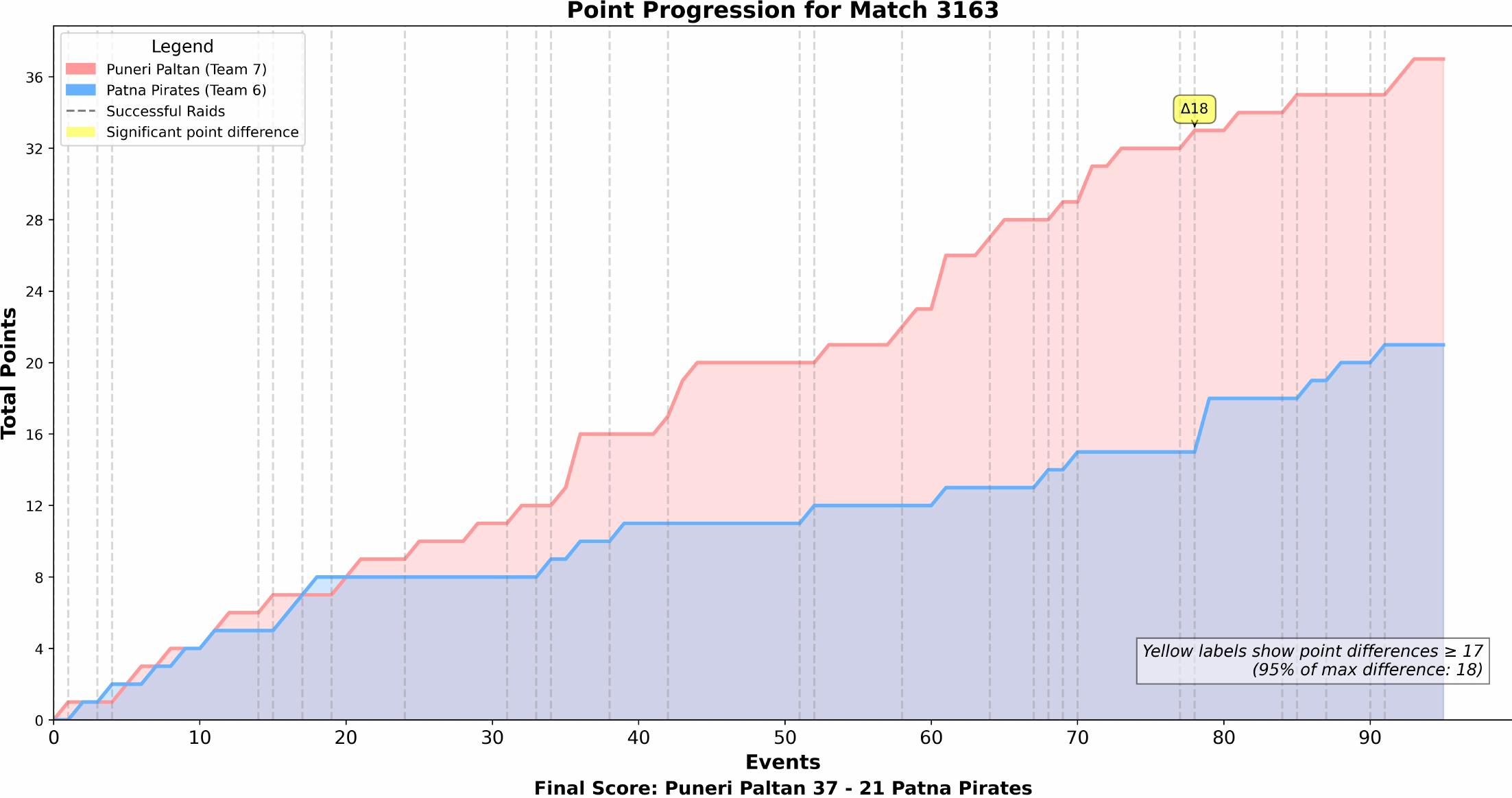}
    \caption{} 
    \label{fig:point-prog}
\end{figure}

We plot the point progression for this 40 minute match with the \texttt{Match\_ID} in Fig~\ref{fig:point-prog}. 

\vspace{-0.2cm}

\begin{lstlisting}
pkl.plot_point_progression(season=10, match_id = 3163)
\end{lstlisting}
\vspace{-0.3cm}

We can clearly see that after an initial period of stagnation between events 20 and 35, Patna Pirates could never recover. This could have been due to several factors, such as a player receiving a red/yellow card, missing a bonus point or the team suffering an all-out. To pinpoint the exact cause we retrieve play-by-play details for the match.

\begin{lstlisting}
roster_desc_pts = pkl.get_team_roster(team_id=4, season=5).sort_values(by='Total Points', ascending=False)

roster_desc_pts.head(10)    
# displaying 10 out of 17 players
\end{lstlisting}



\begin{lstlisting}[style=mystyle_table]
event_no  event             event_text                                  clock   raider_id  defender_id  score
--------------------------------------------------------------------------------------------------------------
21	      Unsuccessful Raid	 Sudhakar M unsuccessful raid	               10:15   4193.0	    NaN	        [9, 8]
22	      Timeout	          None	                                       09:48   NaN	       NaN	        None
23	      Empty Raid	       Aslam Inamdar empty raid	                  09:48   4960.0	    NaN	        [9, 8]
24	      Empty Raid	       Sachin empty raid	                        09:20   757.0	    NaN	        [9, 8]
25	      Successful Raid    Akash Shinde raids successfully	            09:02	  4959.0	    NaN	        [10, 8]
26	      Substitution	    Abinesh Nadarajan comes in for Akash Shinde 08:19	  NaN	       NaN	        None
27	      Empty Raid	       Sachin empty raid	                        08:15	  757.0	    NaN	        [10, 8]
28	      Empty Raid	       Mohit Goyat empty raid	                     07:56	  4022.0	    NaN	        [10, 8]
29	      Unsuccessful Raid	 Sachin unsuccessful raid	                  07:40	  757.0	    4925.0	     [11, 8]
30	      Empty Raid	       Aslam Inamdar empty raid	                  07:07	  4960.0	    NaN	        [11, 8]
31	      Empty Raid	       Babu M empty raid	                        06:39	  726.0	    NaN	        [11, 8]
32	      Successful Raid	 Mohit Goyat raids successfully	            06:24	  4022.0	    NaN	        [12, 8]
33	      Substitution	    Sandeep Kumar comes in for Babu M	         05:46	  NaN	        NaN	        None
34	      Successful Raid	 Sandeep Kumar raids successfully	         05:38	  5282.0	    NaN	        [12, 9]
35	      Successful Raid	 Aslam Inamdar raids successfully	         05:19	  4960.0	    NaN	        [13, 9]
#...Note: Events are sequential, but the clock counts down from 20 minutes for each half of the game
#
#....with 28 more columns 'event_half', 'event_id','raiding_team_id', 'defending_team_id', 'raid_points', 
#       'raid_touch_points', 'raid_bonus_points', 'raid_technical_points', 'raid_all_out_points',
#       'defending_capture_points', 'defending_bonus_points', 'defending_technical_points', 
#       'defending_all_out_points', 'super_raid', 'super_tackle', 'do_or_die', 'super_ten', 'high_five',
#       'review', 'defending_points', 'clock', 'status_id', 'score', 'seq_no',  'defenders', 
#       'created_date', 'player_id', 'substituted_by', 'team_id' and 'substitute_time'
       
      
       
\end{lstlisting}


A quick data query and we have a clear picture of the critical five minute interval where the Patna Pirates were inefficient. They were unable to tackle the offense, while their two consecutive empty raids led to a do-or-die raid, which was unsuccessful and resulted in the concession of an additional point. This resulted in Paltan gaining a lead of 4 points and ultimately winning.

A notable return of this data is the \texttt{raider\_id} and \texttt{defender\_id}'s for each event, which can be used to analyze team dynamics. For example, a raider's performance can be assessed by their success rate against varying numbers of defenders (Fig.~\ref{fig:rvd}) and how their inclusion in the team impacts its overall performance. KabaddiPy enables analysts to easily explore these questions by utilizing a combination of pbp data, player-ids and performance metrics.


\vspace{0.2cm}

For analysts looking to understand the game and the data, season standings can serve as the starting point in their analyses. To demonstrate, we begin by retrieving the season standings for Season 5.


\begin{lstlisting}
print(pkl.get_standings(season=5).head(10))
\end{lstlisting}


\begin{lstlisting}[style=mystyle_table]
Group  Season  Team_Id  Team_Name         League_position  Matches_played  Wins  Lost  Tied
--------------------------------------------------------------------------------------------
B      5       4        Bengal Warriorz   1                22              11    5     6
B      5       6        Patna Pirates     2                22              10    7     5
B      5       30       UP Yoddhas        3                22              8     10    4
B      5       1        Bengaluru Bulls   4                22              8     11    3
B      5       8        Telugu Titans     5                22              7     12    3
B      5       29       Tamil Thalaivas   6                22              6     14    2
A      5       31       Gujarat Giants    1                22              15    4     3
A      5       7        Puneri Paltan     2                22              15    7     0
A      5       28       Haryana Steelers  3                22              13    5     4
A      5       5        U Mumba           4                22              10    12    0
#.....with 5 more columns 'Draws', 'No Result','League_points', 'Score_diff', 'Qualified'
\end{lstlisting}

\noindent Using the \texttt{Team\_Id} obtained, the team level stats can be returned from \texttt{get\_team\_info()} function to see a detailed breakdown of the team stats. We will now examine the stats for the top ranked team in Group B, ``Bengal Warriorz'', with \texttt{Team\_Id} = 4

\begin{lstlisting}
df_rank, df_value, df_per_match, team_raider_skills, team_defender_skills = pkl.get_team_info(season=5, team_id=4)
print(df_value)
\end{lstlisting}

\begin{lstlisting}[style=mystyle_table]
season                                 5.0
-----------------------------------------------------------
team_id                                4
team_name                              Bengal Warriorz
matches_played                         24
team-all-outs-conceded_value           29
team-successful-tackle-percent_value   34.81
team-super-raid_value                  11
...                                    ...
# with 64 more rows across three dataframes each for rank, value, and per match score of team raid,
# successful-raid-percent, dod-raid-points, super-tackles, total_touch_points, total_bonus_points,
# raid-points, successful-raids, total-points-conceded, tackle-points, total-points, successful-tackles, 
# successful-tackles-per-match, all-outs-inflicted, average-raid-points, avg-points-scored and
# average-tackle-points
# and two more dataframes for raider, defender skills with season, skill type, skill name and value.
\end{lstlisting}

\noindent To analyze the gameplay of the Warriorz, we plot the points they scored on the Kabaddi mat (see Fig. 3 and Fig. 4). This data is crucial for opponents, as it highlights key areas of the court where the Warriorz performed well or struggled. By identifying these zones and the top performers within the team, opponents can develop targeted strategies to counter the Warriorz more effectively in future matches.

\begin{lstlisting}
pkl.plot_team_zones(team_id=4, season=5, zone_type='strong')
pkl.plot_team_zones(team_id=4, season=5, zone_type='weak')
\end{lstlisting}

\begin{figure}[h]
    \centering
    \begin{minipage}[b]{0.45\textwidth}
        \centering
        \includegraphics[width=\linewidth]{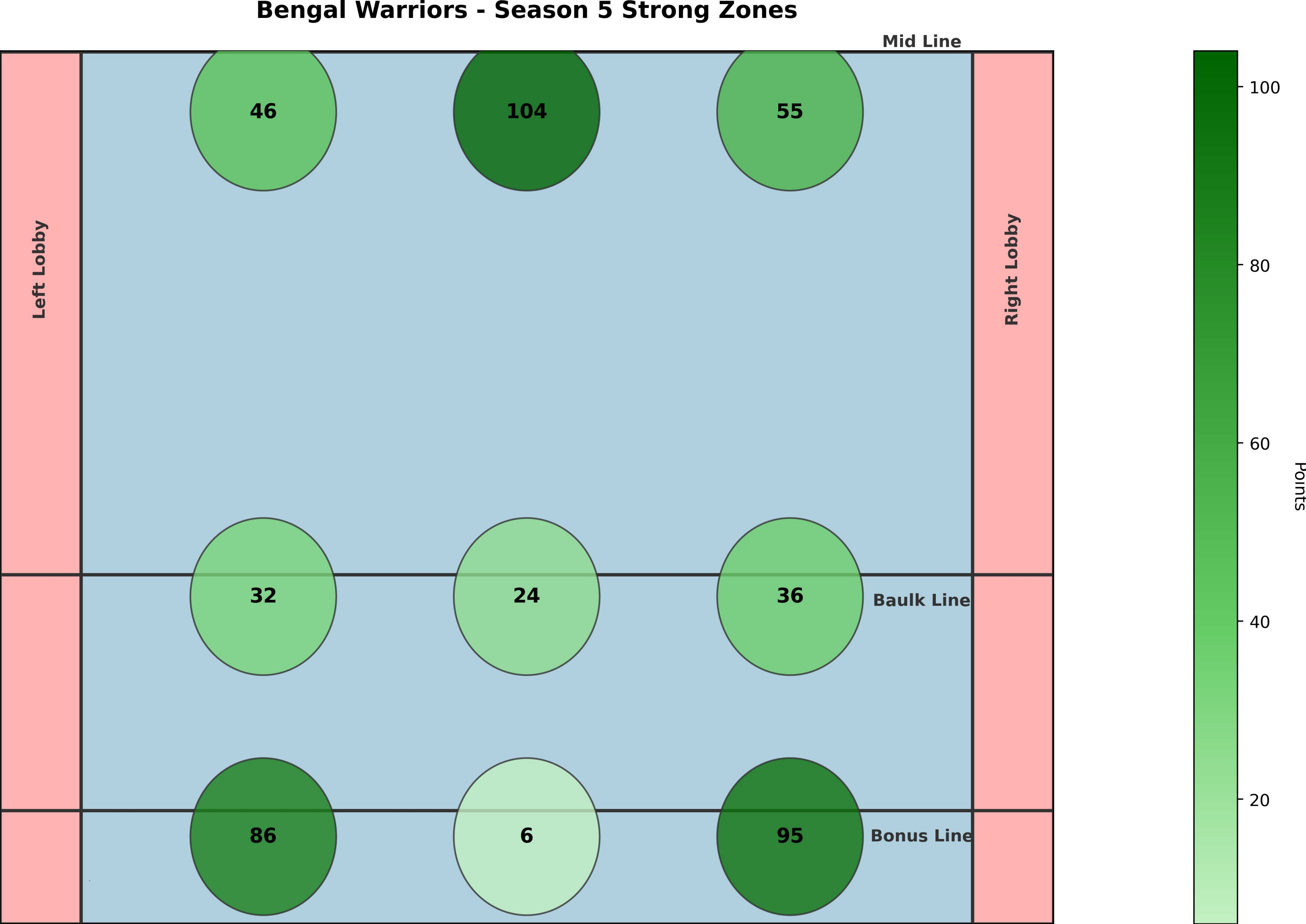}
        \caption{Strong Zones of Bengal Warriorz, Season 5}
        \label{fig:enter-label}
    \end{minipage}
    \hspace{0.05\textwidth} 
    \begin{minipage}[b]{0.45\textwidth}
        \centering
        \includegraphics[width=\linewidth]{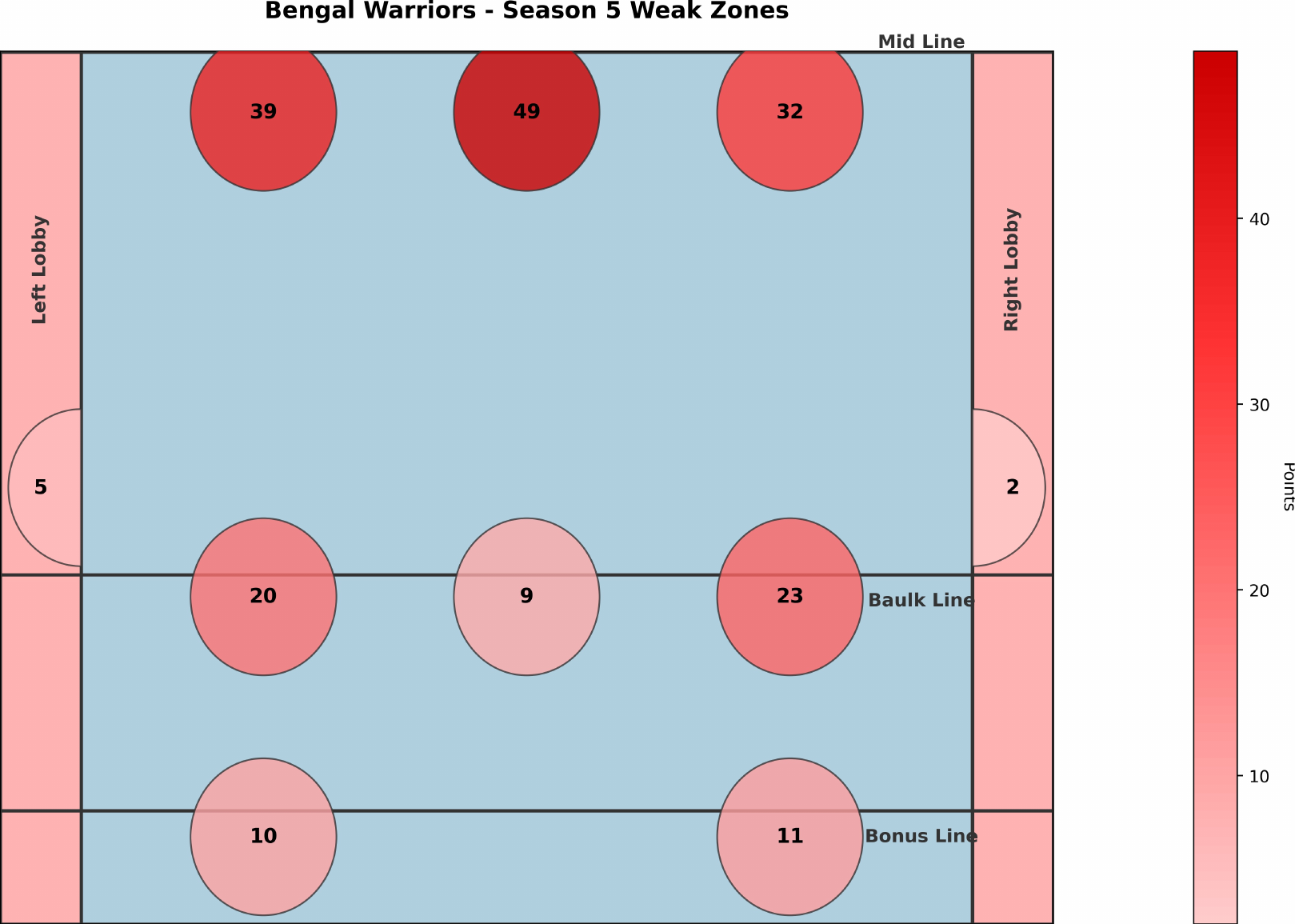}
        \caption{Weak Zones of Bengal Warriorz, Season 5}
        \label{fig:weak}
    \end{minipage}
\end{figure}

\vspace{0.3cm}

\texttt{KabaddiPy} provides an even deeper and more granular view of player-level data. To demonstrate, we call and filter the \texttt{get\_team\_roster()} function to get top five scoring players of Warriorz for Season 5. We plot their strong zones to assess their individual contributions to the team's overall strength.


\begin{lstlisting}
df = pkl.get_team_roster(team_id=4, season=5)

print(df[['Player ID', 'Name','Played Count','Total Points','Team Name', 'Team ID', 'Matches']].sort_values(by='Total Points', ascending=False).head(5))
# using .head() to select 5 players
\end{lstlisting}

We get a list of the top 5 Warriorz players by the total number of points scored in the season.


\begin{lstlisting}[style=mystyle_table]
Player ID   Player Name    Jersey Number Played Count     Total Points    Team Name        Team ID   Matches 
-------------------------------------------------------------------------------------------------------------
143         Maninder Singh  9            21               192             Bengal Warriors   4         24
12          Jang Kun Lee    4            22               89              Bengal Warriors   4         24
211         Deepak Narwal   7            17               87              Bengal Warriors   4         24
322         Surjeet Singh   6            24               79              Bengal Warriors   4         24
160         Ran Singh       13           23               64              Bengal Warriors   4         24
#....with the full team roster having 12 more rows and 8 more columns for 'Captain Count',
#  'Green Card Count', 'Yellow Card Count', 'Red Card Count', 'Starter Count', 'Top Raider Count',
#  'Top Defender Count' and 'Total Matches in Season'
\end{lstlisting}

We use the \texttt{PlayerID} obtained to plot the strong zones for those Warriorz players. (see Fig. 5)

\begin{lstlisting}
plot_player_zones_grid(player_ids=[143, 12, 211, 160], season=5, zone_type='strong', max_cols=2)
\end{lstlisting}

\vspace{-0.1cm}

\begin{figure}[H]
    \centering
    \includegraphics[width=0.9\linewidth]{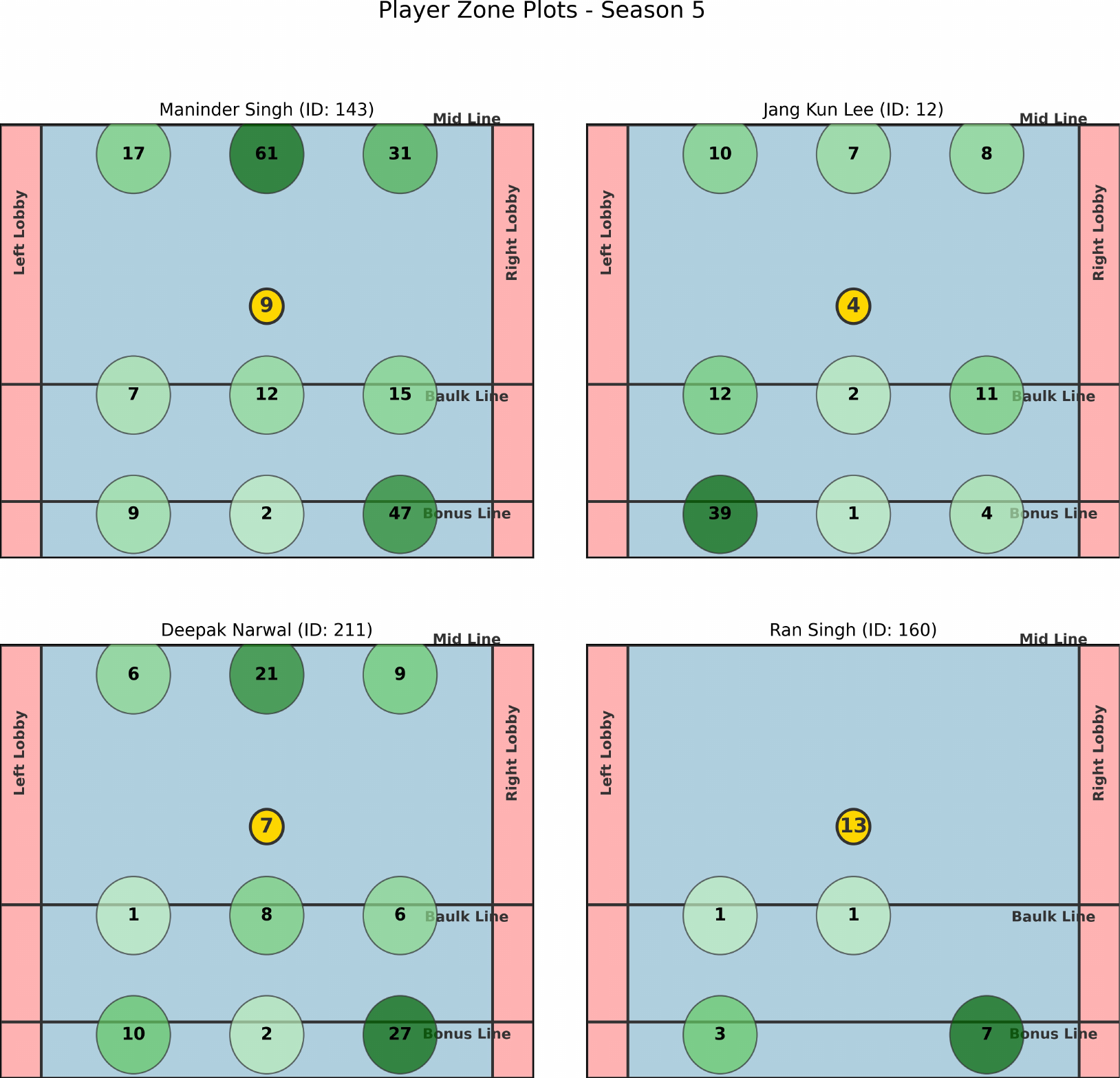}
    \caption{Top Warriorz players strong zones}
    \label{fig:enter-label}
\end{figure}

\vspace{-0.2cm}

\noindent It can be observed that player 211 (Deepak Narwal) has just played 17 matches as compared to 22 matches played by player 12 (Jang Kun Lee) but has scored a comparable number of points. But their strong zones are very different, the former's being the midline and bonus right, while the latter's being the bonus left and the baulk line.

\noindent This facilitates detailed positional analysis and informed team lineup construction while enabling coaches and analysts to optimize player rotations, develop targeted training regimens, and craft opposition-specific strategies to enhance team performance. 


A core Kabaddi dynamic is the performance of a raider against varying numbers of defenders. This metric is critical, as an excellent raid performance against a high number of defenders directly improves a team's winning chances. The function \texttt{get\_player\_rvd()} can be used to analyze this aspect. By providing access to this data across historical seasons, \texttt{KabaddiPy} enables coaches and teams to to identify trends, assess raider efficiency under different defensive pressures and make strategic decisions, such as optimizing substitutions to improve team performance.


We use this function to retrieve and then plot the historical career data for a raider (Maninder Singh, ID 4947) to identify patterns to identify strengths and weaknesses, and develop training strategies accordingly.

\begin{lstlisting}
rvd_data = pkl.get_player_rvd(player_id = 4947)

print(rvd_data[['season', 'player-id', 'Raider Name', 'Team ID','Number of Defenders', 'Total Raids', 'Percentage of Raids', 'Empty Raids Percentage','Successful Raids Percentage']]
\end{lstlisting}

\begin{lstlisting}[style=mystyle_table]
season  player-id  Raider Name      Team_ID  Number_of_Defenders  Total_Raids  Percentage  Successful 
                                                                               of Raids    Raids Percentage
-------------------------------------------------------------------------------------------------------------                                                                                
5       143        Maninder Singh    4        7                   148           40.00%     46.60%
5       143        Maninder Singh    4        5                   51            14.00%     15.70%
5       143        Maninder Singh    4        4                   29            8.00%      34.50%
5       143        Maninder Singh    4        3                   20            5.00%      45.00%
5       143        Maninder Singh    4        2                   26            7.00%      80.80%
...     ...        ...               ...      ...                 ...           ...          ...
9       143        Maninder Singh    4        6                   86            24.00%     66.30%
9       143        Maninder Singh    4        7                   138           38.00%     52.20%
9       143        Maninder Singh    4        1                   1             0%         100.00%
9       143        Maninder Singh    4        2                   28            8%         89.30%
#....with an additional 13 rows and three columns including  'Team Name', 'Empty Raids Percentage'.

\end{lstlisting}

\begin{figure}[h]
    \centering
    \includegraphics[width=0.9\linewidth]{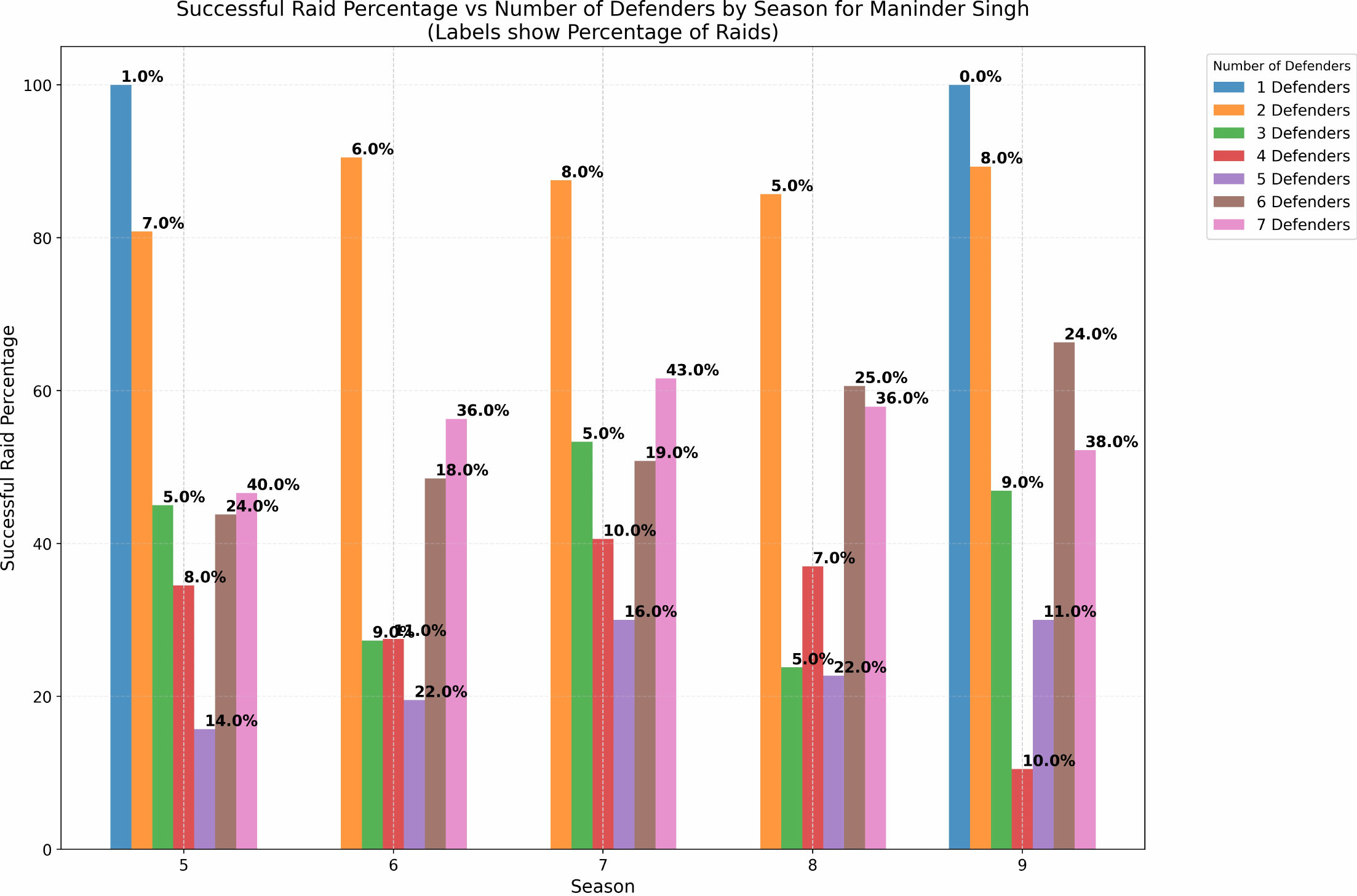}
    \caption{Successful Raid Percentage vs Number of Defenders by Season for Maninder Singh (PlayerID: 143) (Labels show Percentage of that type of Raid)}
    \label{fig:rvd}
\end{figure}

All the \texttt{KabaddiPy} functions, both demonstrated and not, are directly reproducible from the repository. The code demonstrated in the paper can be found \href{https://github.com/kabaddiPy/kabaddiPy/blob/main/examples/0_Reproducing-Paper-Code.ipynb}{here as a Jupyter Notebook}.

\vspace{-0.5cm}

\section{Limitations}



\vspace{-0.5cm}

This module has certain data limitations beyond our control. Specifically, zonal data for seasons 8, 9, and 10 and match breakdown data for season 4 cannot be accessed publicly, potentially affecting the depth of positional and strategic analyses for these metrics. 

Additionally, crucial statistics such as successful raider skills and defender skills as well as raider success rate against particular number of defenders were unavailable for seasons 1 through 4, limiting historical comparisons and trend analyses. To address and avoid these data gaps for future seasons, we propose implementing more robust tracking technologies, and establishing open data-sharing initiatives. 

\section{Conclusion}\label{sec13}

\vspace{-0.35cm}

\texttt{KabaddiPy} was developed as an answer to the lack of consistent and publicly available data for the Pro Kabaddi League (PKL). Both the league and the sport have been rapidly growing in popularity. As the league gains momentum, so does the demand for comprehensive statistics and insights. \texttt{KabaddiPy} delivers reliable data and also promotes reproducible research by ensuring that every dataset and analysis can be consistently replicated.

\vspace{-0.35cm}

\section{What's Next}
\vspace{-0.35cm}

Further steps for \texttt{KabaddiPy} will be to expand its scope to include more rapidly growing international leagues, such as the Kabaddi World Cup and the British Kabaddi League. This will enable cross-league studies and comparisons and offer more insights into the global Kabaddi landscape.

\noindent Although some auction data has been collected, the module's dataset will be expanded, by web scraping, to provide for analysis with respect to this.

\vspace{-0.35cm}
\section{Acknowledgements}
\vspace{-0.35cm}
We thank Alok Pattani, Dean Oliver and Rakshit Naidu for their invaluable feedback on early drafts of this paper. We also thank the owners of the websites for the data.



\bibliography{biblio}

\end{document}